\documentclass[preprint,showpacs,preprintnumbers]{revtex4}
 \usepackage{amsfonts}
 \usepackage{amsmath}
 \usepackage{amssymb}
 \usepackage{graphicx}
 \usepackage{hyperref}
 \makeatletter
 
 \newcommand{\Rmnum}[1]{\expandafter\@slowromancap\romannumeral #1@}
 \makeatother 

\begin{document}
\title{Derivation of exact master equation with stochastic description: Models in quantum optics}
\author{Haifeng Li and Jiushu Shao}\email{jiushu@bnu.edu.cn}
\affiliation{Key Laboratory of Theoretical Computational
Photochemistry, Ministry of Education, College of Chemistry, Beijing
Normal University, Beijing 100875, China}
\begin{abstract}
The methodology of stochastic description for dissipation, a generic
scheme to decouple the interaction between two subsystems, is
applied to the study of dissipative dynamics in quantum optics. It
is shown that the influence of the coupled thermal or vacuum field
on the quantum mode can be exactly represented by the induced
stochastic fields. The quantum mode thereby satisfies a stochastic
differential equation and dissipation effect due to the coupling
with the environment is obtained through statistical averaging.
Within the framework of stochastic description, it is demonstrated
how to derive the master equation for a single optical mode
interacting with the bosonic bath. A numerical algorithm for solving
the master equation in which the coefficients are determined by a
set of integral equations is discussed and a comparison with the
known results is displayed. The derivation of the master equation
for the spontaneous decay of two-state atoms in the vacuum is also
presented.
\end{abstract}
\date{\today}
\maketitle
\section{Introduction}
The best known dissipative dynamics is perhaps the Brownian motion,
which has been greatly explored in theory and well understood since
the pioneering work of Einstein~\cite{Einstein}. Because the
traditional Brownian particle is heavy and its surrounding
environment or bath is at sufficiently high temperature, the motion
of the particle can be accurately described by classical
mechanics~\cite{Langevin}. When the system of interest is very
light, or the temperature of the environment is low, however,
classical mechanics is no longer applicable and one has to invoke
quantum mechanics. Actually, all physical systems intrinsically
follow quantum mechanics and the traditional theory of Brownian
motion should be a classical approximation of the {\em exact}
quantum version. There have been many efforts made to establish a
quantum formulation of dissipative dynamics and the most successful
one is based on the system plus bath model~\cite{Weiss}. Compared to
classical counterpart, the quantum evolution exhibits a unique
feature, that is, coherence. Quantum coherence is a consequence of
the principle of linear superposition and plays indispensable role
in the operation of quantum information devices~\cite{M.A.Nielsen}.
It is also a fundamental issue related to quantum
measurement~\cite{Ford,Schlosshauer,Zurek}. The main purpose of
studies on quantum dissipation is thus to reveal how the environment
affects the time evolution of the quantum system, in particular, the
decoherence effect~\cite{Buchleitner}. \par The classical Brownian
motion is generally described by a generalized Langevin equation in
which the random force is induced by the thermal
bath~\cite{Langevin, Weiss}. One naturally wonders whether the
impact of the bath can be defined as a {\em classical} random field.
Kubo was the first to propose the stochastic Liouville equation for
quantum dissipative systems, but his formulation is
phenomenological~\cite{Kubo}. Given the system plus bath model, the
key issue is then to {\em acquire} the equation of motion of the
system, in which the dissipation effect due to the bath is exactly
taken into account and the explicit dynamics of the bath does not
show up. In other words, one aims at finding the effective motion of
the system in its own physical space instead of inspecting the every
detail of the dynamics for the whole system. To this end, several
theoretical frameworks including the projection operator
technique~\cite{Nakajima, Zwanzig}, the influence functional
method~\cite{Feynman, Hibbs, Caldeira}, and quantum Langevin
equation approach~\cite{Benguria,Ford1,Risken} were suggested and
applied to a wide range of physical systems. Of course, all of these
methods have their own pros and cons in practice. We have recently
put forward a stochastic theory for dissipative systems, in which
the interaction with the bath is rigorously mapped into stochastic
fields during the evolution of the system~\cite{J.Shao}. In this
formulation the system is subjected to complex stochastic fields
induced by the bath~\cite{J.Shao}. For comparison, in Kubo's
stochastic Liouville equation, there is only a real stochastic field
due to the bath. For specific dissipation, we have proposed the
hierarchical equation of motion approach in terms of the stochastic
formulation~\cite{Shao1, Tanimura}, which has become an efficient,
deterministic numerical technique of many applications~\cite{Shao2}.
Combining the stochastic and the deterministic methods, we were able
to solve the dynamics of the two-state system strongly coupled to a
bosonic bath~\cite{Shao3}. Besides the trophy in numerical
simulations, it has been shown that the stochastic formulation
offers a convenient, systematic procedure for theoretical analysis,
say the derivation of the master equation for linear
systems~\cite{Li}. When the existence of master equation is
warranted, its derivation and solution should be the central task in
quantum dissipative dynamics. This paper, as a continuation of the
previous one~\cite{Li}, addresses the dissipative dynamics in
quantum optics. We will apply the stochastic formulation to derive
the well-known master equations for two models. One is a single mode
perturbed by a thermal bath~\cite{Zhang} and the other is the
spontaneous decay: the two-state atom in the
vacuum~\cite{J.Shao,Breure,Garraway, H.P.Breuer, Vacchini,
Smirne,Strunz}. \par As discussed in literature, the fluctuations of
the thermal field are a major source of dissipation in quantum
optics, which damage coherence of the system~\cite{Carmichael}.
Again, because quantum optical devices operate when decoherence is
negligible, to realize optimal functioning, it is sometimes
necessary to design a scheme to control
dissipation~\cite{Breure,Carmichael,Wolf,Glauber}. This entails a
clear revelation of underlying dissipative dynamics. It is no doubt
that exactly solvable models may provide fundamental understanding
in this respect and are always desired. In the previous paper we
have shown how to employ the stochastic description of quantum
dissipation to work out the master equation~\cite{Li}. The harmonic
oscillator coupled to the Caldeira-Leggett heat bath is used as an
example. In this paper the thermal field as well as the vacuum is
considered to be the heat bath. There are two kinds of interaction
between the system and the bath, one corresponding to the absorption
and the other the emission, of a photon
energy~\cite{Zhang,Gardiner,Walls,Pierre}. Although these models can
formally be regarded as a result of rotating-wave approximation
(RWA) imposed on the Caldeira-Leggett model, we will avoid the
discussion on the validity of the approximation~\cite{Fleming}. \par
The first model we shall consider is essentially the dissipative
harmonic oscillator within RWA. Its master equation was derived by
Xiong {\em et al.}~\cite{Zhang}, resorting to the influence
functional approach developed by Feynman and Vernon~\cite{Feynman,
Hibbs}. As the authors observed, when the coupling between the
system and the bath becomes strong, the dissipative dynamics will
change dramatically because of the non-Markovian memory
effect~\cite{Zhang}. In a recent paper~\cite{Tan}, Tan and Zhang
used the same method to discuss the consequence of initial
system-bath correlation. The second model describes the spontaneous
decay of two-state atoms in vacuum, which is exactly solvable. It
has been frequently used as benchmark calculations in quantum
optics. Indeed, this model is so well-known that diversified methods
have been used to derive its master
equation~\cite{Breure,Garraway,H.P.Breuer,J.Shao,Vacchini,Smirne,Strunz}.

\par The paper is organized as follows.
In Sec.~\Rmnum{2} we recapitulate the stochastic formulation for
quantum dissipation. In Sec.~\Rmnum{3} we apply the scheme to a
single-mode cavity system coupled to a thermal field and derive its
exact master equation. In Sec.~\Rmnum{4} the obtained master
equation is shown to be equivalent to the result in
Ref.~\cite{Zhang} and some remarks on the numerical implementation
are given. In Sec.~\Rmnum{5} the master equation of the same system
subjected to a driving external field is derived. In Sec.~\Rmnum{6}
we show how to derive the master equation of a two-state atom
coupled to the vacuum field. We present our conclusions in
Sec.~\Rmnum{7}.
\section{Theory}
To study the dissipative dynamics of a quantum mode in an optical
cavity, we start with an arbitrary cavity system coupled to a
thermal field consisting of infinite number of harmonic oscillators.
The Hamiltonian of the entire system assumes
 \begin{equation}
\hat{H}
=\hat{H}_{s}+\sum_{j}\hbar\omega_{j}b_{j}^{\dagger}b_{j}+\sum_{j}\hbar\left(
c_{j}\hat{f}_{1}b_{j}^{\dagger}+c_{j}\hat{f}_{2}b_{j}\right),
 \end{equation}
where the first term on the right-hand side is the Hamiltonian of
the cavity mode, the second term is the Hamiltonian of the thermal
bath, and the last two terms define the interaction between the
system and the bath. Here~$\hat{f}_{1}$~and~$\hat{f}_{2}$~are
operators for the system and they are a hermitian
pair,~$\hat{f}_{1}=\hat{f}^{+}_{2}$. Note that the two interaction
terms can be interpreted as emitting and absorbing a quantum phonon
or photon by the bath. The model will be the Caldeira-Leggett type
when the interaction is of the form~$\sum_{j}\hbar
c_{j}\left(\hat{f}_{1}+\hat{f}_{2}\right)\left(b_{j}^{\dagger}+b_{j}\right)$.
As shown in the previous papers~\cite{J.Shao}, the dissipative
dynamics can be described by a stochastic formulation in which the
system evolves in the stochastic fields induced by the bath and the
statistical average of the random density matrix is nothing but the
reduced density matrix. For the model we consider, the random
density matrix satisfies
\begin{align}
i\hbar
d\rho_{s}(t)=&\left[\hat{H}_{s}+\sum_{k=1}^{2}\bar{g}_{k}(t)\hat{f}_{k},\rho_{s}\right]dt
+\frac{\sqrt{\hbar}}{2}\sum_{k=1}^{2}\left[\hat{f}_{k},\rho_{s}\right]dW_{1k}
+i\frac{\sqrt{\hbar}}{2}\sum_{k=1}^{2}\left\{\hat{f}_{k},\rho_{s}\right\}dW^{*}_{2k},
\end{align}
where the bath-induced stochastic fields are given by~\begin{align}
\bar{g}_{1}(t)=\sum_{j}\frac{\textup{Tr}_{b}\left\{\hbar
c_{j}b^{\dagger}_{j}\rho_{b}(t)\right\}}{\textup{Tr}_{b}\left\{\rho_{b}(t)\right\}}
\intertext{and}
\bar{g}_{2}(t)=\sum_{j}\frac{\textup{Tr}_{b}\left\{\hbar
c_{j}b_{j}\rho_{b}(t)\right\}}{\textup{Tr}_{b}\left\{\rho_{b}(t)\right\}}.\end{align}
Here, introduced are the complex Wiener processes
$W_{1k}(t)=\int_{0}^{t}dt^{\prime}\left[\nu_{1k}(t^{\prime})+i\nu_{4k}(t^{\prime})\right]$~and
$W_{2k}(t)=\int_{0}^{t}dt^{\prime}\left[\nu_{2k}(t^{\prime})+i\nu_{3k}(t^{\prime})\right]$,
where~$\nu_{nk}(t)~(n=1-4)$~are independent Gaussian white noises
with zero mean and delta function correlation. This is the main
result of the stochastic formulation and will be the working
formula. To use it, of course, we need to first
calculate~$\bar{g}_{1}(t)$~and~$\bar{g}_{2}(t)$. In this
formulation, $\bar{g}_{1}(t)$~and~$\bar{g}_{2}(t)$~can be determined
by the evolution of the bath,
\begin{align}
i\hbar
d\rho_{b}=&\sum_{j}\left[\hbar\omega_{j}b_{j}^{\dagger}b_{j},\rho_{b}\right]dt+\frac{\hbar\sqrt{\hbar}}{2}\sum_{j}
c_{j}\left[b^{\dagger}_{j},\rho_{b}\right]dW_{21}+\frac{\hbar
\sqrt{\hbar}}{2}\sum_{j}c_{j}\left[b_{j},\rho_{b}\right]dW_{22}\nonumber\\
&+i\frac{\hbar
\sqrt{\hbar}}{2}\sum_{j}c_{j}\left\{b^{\dagger}_{j},\rho_{b}\right\}dW^{*}_{11}+i\frac{
\hbar\sqrt{\hbar}}{2}\sum_{j}c_{j}\left\{b_{j},\rho_{b}\right\}dW^{*}_{12}.
\end{align}
The formal solution of~$\rho_{b}(t)$~can be written as
\begin{equation}
\rho_{b}(t)=u_{1}(t,0)\rho_{b}(0)u_{2}(0,t),
\end{equation}
where~$u_{1,2}(t,0)$~are the forward and backward propagators
dictated by
\begin{align*}
\hat{h}_{1}(t)=&\sum_{j}\hbar\omega_{j}b_{j}^{\dagger}b_{j}+\frac{\hbar\sqrt{\hbar}}{2}\sum_{j}c_{j}b_{j}^{\dagger}\eta_{11}(t)+\frac{\hbar\sqrt{\hbar}}{2}\sum_{j}c_{j}b_{j}\eta_{12}(t)
\end{align*}
and
\begin{align*}
\hat{h}_{2}(t)=&\sum_{j}\hbar\omega_{j}b_{j}^{\dagger}b_{j}+\frac{\hbar\sqrt{\hbar}}{2}\sum_{j}c_{j}b_{j}^{\dagger}\eta_{21}(t)+\frac{\hbar\sqrt{\hbar}}{2}\sum_{j}c_{j}b_{j}\eta_{22}(t).
\end{align*}
with
\begin{align*}
\eta_{11}(t)&=\nu_{21}(t)+i\nu_{31}(t)+i\nu_{11}(t)+\nu_{41}(t),\\
\eta_{12}(t)&=\nu_{22}(t)+i\nu_{32}(t)+i\nu_{12}(t)+\nu_{42}(t),\\
\eta_{21}(t)&=\nu_{21}(t)+i\nu_{31}(t)-i\nu_{11}(t)-\nu_{41}(t),\\
\eta_{22}(t)&=\nu_{22}(t)+i\nu_{32}(t)-i\nu_{12}(t)-\nu_{42}(t),
\end{align*}
being complex white noises. Because the bath modes are independent,
the propagator of the bath is a product of the individual ones,
namely,~$u_{1}(t,0)=\prod_{j}u_{j,1}(t,0)$~and~$u_{2}(0,t)=\prod_{j}u_{j,2}(0,t)$.
\par
As illustrated in the previous paper and other
references~\cite{Li,Louisell}, the propagator for each bath mode can
feasibly be obtained upon using the interaction representation. As a
result, the forward propagator~$u_{1,j}(t,0)$~reads
\begin{equation}
u_{j,1}(t,0)=C_{j,10}(t)e^{C_{j,11}(t)b_{j}}e^{C_{j,12}(t)b_{j}^{\dagger}}u_{j,0}(t,0),
\end{equation}
where~$u_{j,0}(t,0)$~is the propagator of the undriven harmonic
oscillator described
by~$h_{j,0}=\hbar\omega_{j}b_{j}^{\dagger}b_{j}$, which is well
known~\cite{Feynman, Hibbs,Dittrich,Kleinert,R.P.Feynman}, and
\begin{align*}
C_{j,10}(t)&=\exp\left[\frac{\hbar}{4}c^{2}_{j}\int_{0}^{t}dt_{1}\int_{0}^{t_{1}}dt_{2}\eta_{11}(t_{1})\eta_{12}(t_{2})e^{i\omega_{j}(t_{1}-t_{2})}\right],\\
C_{j,11}(t)&=-i\frac{\sqrt{\hbar}}{2}c_{j}\int_{0}^{t}dt_{1}\eta_{12}(t_{1})e^{i\omega_{j}(t-t_{1})},\\
C_{j,12}(t)&=-i\frac{\sqrt{\hbar}}{2}c_{j}\int_{0}^{t}dt_{1}\eta_{11}(t_{1})e^{-i\omega_{j}(t-t_{1})}.
\end{align*}
Similarly, the backward propagator~$u_{j,2}(0,t)$~is
\begin{equation}
u_{j,2}(0,t)=C_{j,20}(t)u_{j,0}(0,t)e^{C_{j,22}(t)b_{j}^{\dagger}}e^{C_{j,21}(t)b_{j}},
\end{equation}
where
\begin{align*}
C_{j,20}(t)&=\exp\left[-\frac{\hbar}{4} c^{2}_{j}\int_{0}^{t}dt_{1}\int_{0}^{t_{1}}dt_{2}\eta_{21}(t_{1})\eta_{22}(t_{2})e^{i\omega_{j}(t_{1}-t_{2})}\right],\\
C_{j,21}(t)&=i\frac{\sqrt{\hbar}}{2}c_{j}\int_{0}^{t}dt_{1}\eta_{22}(t_{1})e^{i\omega_{j}(t-t_{1})},\\
C_{j,22}(t)&=i\frac{\sqrt{\hbar}}{2}c_{j}\int_{0}^{t}dt_{1}\eta_{21}(t_{1})e^{-i\omega_{j}(t-t_{1})}.
\end{align*}
Assume that the thermal field is initially in thermal equilibrium,
\begin{equation}
\rho_{b}(0)=\frac{1}{Z_{b}}e^{-\beta\hat{H}_{b}}=\frac{1}{Z_{b}}\prod_{j}e^{-\beta
h_{j,0}},
\end{equation}
where~$Z_{b}=\textup{Tr}_{b}\left\{e^{-\beta\hat{H}_{b}}\right\}$~is
the partition function. Inserting together with Eqs.~(7) and (8)
into Eq.~(6), carrying some operator algebra and rearranging, we
obtain \begin{align} \rho_{b}(t)=\prod_{j}F_{j}(t)
\exp\left\{\left[C_{j,12}(t)+C_{j,22}(t)e^{-\beta\hbar\omega_{j}}\right]b_{j}^{\dagger}\right\}
\exp\left\{\left[C_{j,11}(t)+C_{j,21}(t)e^{\beta\hbar\omega_{j}}\right]b_{j}\right\}\exp\left(-\beta\hat{H}_{b}\right),\end{align}
where
\begin{equation*}
F_{j}(t)=\frac{1}{Z_{b}}C_{j,10}(t)C_{j,20}(t)\exp\left\{C_{j,11}(t)\left[C_{j,12}(t)+C_{j,22}(t)e^{-\beta\hbar\omega_{j}}\right]\right\}.
\end{equation*}
Then the bath-induced fields defined by Eqs.~(3) and (4) can be
worked out,
\begin{equation}
\bar{g}_{1}(t)=\frac{i\hbar\sqrt{\hbar}}{2}
\int_{0}^{t}dt^{\prime}\left\{\alpha_{1}(t-t^{\prime})\left[\nu_{22}(t^{\prime})
+i\nu_{32}(t^{\prime})\right]-\alpha_{2}(t-t^{\prime})\left[i\nu_{12}(t^{\prime})+\nu_{42}(t^{\prime})\right]\right\}
\end{equation}
and
\begin{equation}
\bar{g}_{2}(t)=-\frac{i\hbar\sqrt{\hbar}}{2}\int_{0}^{t}dt^{\prime}\{\alpha^{*}_{1}(t-t^{\prime})[\nu_{21}(t^{\prime})+i\nu_{31}(t^{\prime})]
+\alpha^{*}_{2}(t-t^{\prime})[i\nu_{11}(t^{\prime})+\nu_{41}(t^{\prime})]\},
\end{equation}
where $\alpha_{1}(t)$~and~$\alpha_{2}(t)$~are response functions
given by
\begin{align}
\alpha_{1}(t)&=
\int_{0}^{\infty}d\omega J(\omega)e^{i\omega t}\\
\intertext{and} \alpha_{2}(t)&=\int_{0}^{\infty}d\omega
J(\omega)\coth\left(\frac{\beta}{2}\hbar\omega\right)e^{i\omega t},
\end{align}

$J(\omega)$~being the spectral density function
\begin{equation}
J(\omega)=\sum_{j}c_{j}^{2}\delta(\omega_{j}-\omega).
\end{equation}
We like to stress that~$J(\omega)$~completely captures the influence
of the thermal field on the system. With the
known~$\bar{g}_{1,2}(t)$, Eq.~(2) is a closed stochastic
differential equation. That is, one can generate the required
stochastic fields through white noises and solve Eq.~(2) for a given
initial condition~$\rho_{s}(0)$. The reduced density
matrix~$\tilde{\rho}_{s}(t)$~is of course the mathematical
expectation of random density matrix~$\rho_{s}(t)$, namely,
$\tilde{\rho}_{s}(t)=M\left\{\rho_{s}(t)\right\}$. We can also try
to find the equation of motion for~$\tilde{\rho}_{s}(t)$~from
Eq.~(2). To this end, we take stochastic averaging to obtain
\begin{align}
i\hbar {d\tilde{\rho}_{s} \over
dt}=&\left[\hat{H}_{s},\tilde{\rho}_{s}\right]+\frac{i\hbar\sqrt{\hbar}}{2}\left[\hat{f}_{1},\int_{0}^{t}dt^{\prime}\left[\alpha_{1}(t-t^{\prime})
\hat{O}_{s,11}(t,t^{\prime})
-\alpha_{2}(t-t^{\prime})\hat{O}_{s,12}(t,t^{\prime})\right]\right]\nonumber\\
&-\frac{i\hbar\sqrt{\hbar}}{2}\left[\hat{f}_{2},\int_{0}^{t}dt^{\prime}\left[\alpha^{*}_{1}(t-t^{\prime})
\hat{O}_{s,21}(t,t^{\prime})+\alpha^{*}_{2}(t-t^{\prime})\hat{O}_{s,22}(t,t^{\prime})\right]\right],
\end{align}
where the dissipative operators are
\begin{align}
\hat{O}_{s,11}(t,t^{\prime})=&M\left\{\frac{\delta\rho_{s}(t)}{\delta\nu_{22}(t^{\prime})}+i\frac{\delta\rho_{s}(t)}{\delta\nu_{32}(t^{\prime})}\right\},
\\
\hat{O}_{s,12}(t,t^{\prime})=&M\left\{i\frac{\delta\rho_{s}(t)}{\delta\nu_{12}(t^{\prime})}+\frac{\delta\rho_{s}(t)}{\delta\nu_{42}(t^{\prime})}\right\}
,\\
\hat{O}_{s,21}(t,t^{\prime})=&M\left\{\frac{\delta\rho_{s}(t)}{\delta\nu_{21}(t^{\prime})}+i\frac{\delta\rho_{s}(t)}{\delta\nu_{31}(t^{\prime})}\right\},\\
\hat{O}_{s,22}(t,t^{\prime})=&M\left\{i\frac{\delta\rho_{s}(t)}{\delta\nu_{11}(t^{\prime})}+\frac{\delta\rho_{s}(t)}{\delta\nu_{41}(t^{\prime})}\right\}.
\end{align}

In the above derivation the nonanticipating property
of~$\rho_{s}(t)$, namely,
$M\left\{\rho_{s}(t)dW_{m,k}(t)\right\}=0~(m~$=$~1,2)$, and the
Furutsu-Novikov theorem~\cite{Novikov}, that is, $
M\left\{\nu(t^{\prime})F\left[\nu\right]\right\}=M\left\{\delta
F[\nu]/\delta\nu(t^{\prime})\right\}$~for a white noise~$\nu(t)$~and
its arbitrary functional~$F[\nu]$, are used. As shown in
Ref.~\cite{J.Shao}, the formal solution of
Liouville-like~Eq.~(2)~can be written as
\begin{equation}
\rho_{s}(t)=U_{1}(t,0)\rho_{s}(0)U_{2}(0,t),
\end{equation}
where~$U_{1}(t,0)$~is the forward propagator associated with the
Hamiltonian
\begin{equation}
\hat{H}_{1}(t)=\hat{H}_{s}+\xi_{11}(t)\hat{f}_{1}+\xi_{12}(t)\hat{f}_{2}
\end{equation}
while~$U_{2}(0,t)$~is the backward propagator associated with the
Hamiltonian
\begin{equation}
\hat{H}_{2}(t)=\hat{H}_{s}+\xi_{21}(t)\hat{f}_{1}+\xi_{22}(t)\hat{f}_{2}
\end{equation}
with
\begin{align*}
\xi_{11}(t)&=\bar{g}_{1}(t)+i\frac{\sqrt{\hbar}}{2}\eta^{*}_{11}(t),\\
\xi_{12}(t)&=\bar{g}_{2}(t)+i\frac{\sqrt{\hbar}}{2}\eta^{*}_{12}(t),\\
\xi_{21}(t)&=\bar{g}_{1}(t)-i\frac{\sqrt{\hbar}}{2}\eta^{*}_{21}(t),\\
\xi_{22}(t)&=\bar{g}_{2}(t)-i\frac{\sqrt{\hbar}}{2}\eta^{*}_{22}(t).
\end{align*}
Following Ref~\cite{J.Shao}, we calculate the functional derivatives
to obtain {\em formal} solutions of the dissipative operators,
\begin{align}
\hat{O}_{s,11}(t,t^{\prime})&=\frac{1}{\sqrt{\hbar}}
M\left\{U_{1}(t,t^{\prime})\hat{f}_{2}U_{1}(t^{\prime},t)\rho_{s}(t)+\rho_{s}(t)U_{2}(t,t^{\prime})\hat{f}_{2}U_{2}(t^{\prime},t)\right\},\\
\hat{O}_{s,12}(t,t^{\prime})&=\frac{1}{\sqrt{\hbar}}M\left
\{U_{1}(t,t^{\prime})\hat{f}_{2}U_{1}(t^{\prime},t)\rho_{s}(t)-\rho_{s}(t)U_{2}(t,t^{\prime})\hat{f}_{2}U_{2}(t^{\prime},t)\right\},\\
\hat{O}_{s,21}(t,t^{\prime})&=\frac{1}{\sqrt{\hbar}}M\left
\{U_{1}(t,t^{\prime})\hat{f}_{1}U_{1}(t^{\prime},t)\rho_{s}(t)+\rho_{s}(t)U_{2}(t,t^{\prime})\hat{f}_{1}U_{2}(t^{\prime},t)\right\},\\
\hat{O}_{s,22}(t,t^{\prime})&=\frac{1}{\sqrt{\hbar}}
M\left\{U_{1}(t,t^{\prime})\hat{f}_{1}U_{1}(t^{\prime},t)\rho_{s}(t)-\rho_{s}(t)U_{2}(t,t^{\prime})\hat{f}_{1}U_{2}(t^{\prime},t)\right\}.
\end{align}
When these operators can be expressed in terms of the reduced
density operator~$\tilde{\rho}_{s}(t)$~and other operators of the
system, then Eq.~(16) becomes a master equation. In the following
section, we show that a dissipative single-mode optical cavity is
indeed the case.
\section{Master equation for dissipative single-mode optical cavity}
Let us consider a single-mode cavity,
$\hat{H}_{s}=\hbar\omega_{0}a^{\dagger}a$, with the coupling
operators~$\hat{f}_{1}=a$~and~$\hat{f}_{2}=a^{\dagger}$. Therefore,
the dynamics of the random cavity are determined by the forward and
backward Hamiltonians
\begin{align}
\hat{H}_{1}(t)=\hbar\omega_{0}a^{\dagger}a+\xi_{11}(t)a+\xi_{12}(t)a^{\dagger}
\intertext{and}
\hat{H}_{2}(t)=\hbar\omega_{0}a^{\dagger}a+\xi_{21}(t)a+\xi_{22}(t)a^{\dagger}.
\end{align}
These Hamiltonians are exactly solvable and one readily finds
\begin{align*}
U_{1}(t,t^{\prime})aU_{1}(t^{\prime},t)=&ae^{i\omega_{0}(t-t^{\prime})}+\frac{i}{\hbar}\int_{t^{\prime}}^{t}dt_{1}
\xi_{12}(t_{1})e^{i\omega_{0}(t_{1}-t^{\prime})},\\
U_{2}(t,t^{\prime})aU_{2}(t^{\prime},t)=&ae^{i\omega_{0}(t-t^{\prime})}+\frac{i}{\hbar}\int_{t^{\prime}}^{t}dt_{1}
\xi_{22}(t_{1})e^{i\omega_{0}(t_{1}-t^{\prime})},\\
U_{1}(t,t^{\prime})a^{\dagger}U_{1}(t^{\prime},t)=&a^{\dagger}e^{-i\omega_{0}(t-t^{\prime})}-\frac{i}{\hbar}\int_{t^{\prime}}^{t}dt_{1}
\xi_{11}(t_{1})e^{-i\omega_{0}(t_{1}-t^{\prime})},\\
U_{2}(t,t^{\prime})a^{\dagger}U_{2}(t^{\prime},t)=&a^{\dagger}e^{-i\omega_{0}(t-t^{\prime})}-\frac{i}{\hbar}\int_{t^{\prime}}^{t}dt_{1}
\xi_{21}(t_{1})e^{-i\omega_{0}(t_{1}-t^{\prime})}.
\end{align*}
Inserting into Eqs.~(24)$-$(27) and taking statistical averaging, we
obtain
\begin{align}
\hat{O}_{s,11}(t,t^{\prime})=&\frac{1}{\sqrt{\hbar}}\left\{a^{\dagger},\tilde{\rho}_{s}(t)\right\}e^{-i\omega_{0}
(t-t^{\prime})}+\int_{t^{\prime}}^{t}dt_{1}\int_{0}^{t_{1}}dt_{2}e^{-i\omega_{0}(t_{1}-t^{\prime})}\alpha_{1}(t_{1}-t_{2})\hat{O}_{s,11}(t,t_{2})
\nonumber\\&-\int_{t^{\prime}}^{t}dt_{1}\int_{0}^{t}dt_{2}e^{-i\omega_{0}(t_{1}-t^{\prime})}
\alpha_{2}(t_{1}-t_{2})\hat{O}_{s,12}(t,t_{2}),\\
\hat{O}_{s,12}(t,t^{\prime})=&\frac{1}{\sqrt{\hbar}}\left[a^{\dagger},\tilde{\rho}_{s}(t)\right]e^{-i\omega_{0}
(t-t^{\prime})}-\int_{t^{\prime}}^{t}dt_{1}
\int_{t_{1}}^{t}dt_{2}e^{-i\omega_{0}(t_{1}-t^{\prime})}\alpha^{*}_{1}(t_{2}-t_{1})\hat{O}_{s,12}(t,t_{2}),\\
\hat{O}_{s,21}(t,t^{\prime})=&\frac{1}{\sqrt{\hbar}}
\left\{a,\tilde{\rho}_{s}(t)\right\}e^{i\omega_{0}
(t-t^{\prime})}+\int_{t^{\prime}}^{t}dt_{1}\int_{0}^{t_{1}}dt_{2}e^{i\omega_{0}(t_{1}-t^{\prime})}
\alpha^{*}_{1}(t_{1}-t_{2})\hat{O}_{s,21}(t,t_{2})
\nonumber\\&+\int_{t^{\prime}}^{t}dt_{1}\int_{0}^{t}dt_{2}e^{i\omega_{0}(t_{1}-t^{\prime})}\alpha_{2}(t_{2}-t_{1})
\hat{O}_{s,22}(t,t_{2}),\\
\hat{O}_{s,22}(t,t^{\prime})=&\frac{1}{\sqrt{\hbar}}\left[a,\tilde{\rho}_{s}(t)\right]e^{i\omega_{0}
(t-t^{\prime})}-\int_{t^{\prime}}^{t}dt_{1}
\int_{t_{1}}^{t}dt_{2}e^{i\omega_{0}(t_{1}-t^{\prime})}\alpha_{1}(t_{2}-t_{1})\hat{O}_{s,22}(t,t_{2}).
\end{align}
In the above derivation, the following functional derivatives with
respective white noises are used,
\begin{align*}
M\left\{\frac{\delta\rho_{s}(t)}{\delta\nu_{11}(t^{\prime})}+i\frac{\delta\rho_{s}(t)}{\delta\nu_{41}(t^{\prime})}\right\}
=&-i\hbar\int_{t^{\prime}}^{t}dt_{1}
\alpha_{2}^{*}(t_{1}-t^{\prime})\hat{O}_{s,12}(t,t_{1}),\\
M\left\{i\frac{\delta\rho_{s}(t)}{\delta\nu_{21}(t^{\prime})}+\frac{\delta\rho_{s}(t)}{\delta\nu_{31}(t^{\prime})}\right\}=&
-i\hbar\int_{t^{\prime}}^{t}dt_{1}\alpha_{1}^{*}(t_{1}-t^{\prime})\hat{O}_{s,12}(t,t_{1}),\\
M\left\{\frac{\delta\rho_{s}(t)}{\delta\nu_{12}(t^{\prime})}+i\frac{\delta\rho_{s}(t)}{\delta\nu_{42}(t^{\prime})}\right\}
=&-i\hbar\int_{t^{\prime}}^{t}dt_{1}\alpha_{2}(t_{1}-t^{\prime})\hat{O}_{s,22}(t,t_{1}),\\
M\left\{i\frac{\delta\rho_{s}(t)}{\delta\nu_{22}(t^{\prime})}+\frac{\delta\rho_{s}(t)}{\delta\nu_{32}(t^{\prime})}\right\}
=&i\hbar\int_{t^{\prime}}^{t}dt_{1}\alpha_{1}(t_{1}-t^{\prime})\hat{O}_{s,22}(t,t_{1}),
\end{align*}
which can be found directly through the solution of~$\rho_{s}(t)$.
Note
that~$\hat{O}_{s,11}^{\dagger}(t,t^{\prime})=\hat{O}_{s,21}(t,t^{\prime})$~and
$\hat{O}_{s,12}^{\dagger}(t,t^{\prime})=-\hat{O}_{s,22}(t,t^{\prime})$.
By iteration, one can show that the dissipative operators should
assume the following forms,
\begin{align}
\hat{O}_{s,11}(t,t^{\prime})=&x_{11}(t,t^{\prime})\left\{a^{\dagger},\tilde{\rho}_{s}(t)\right\}
+x_{12}(t,t^{\prime})\left[a^{\dagger},\tilde{\rho}_{s}(t)\right],\\
\hat{O}_{s,12}(t,t^{\prime})=&x_{21}(t,t^{\prime})\left[a^{\dagger},\tilde{\rho}_{s}(t)\right],\\
\hat{O}_{s,21}(t,t^{\prime})=&x^{*}_{11}(t,t^{\prime})\left\{a,\tilde{\rho}_{s}(t)\right\}-x^{*}_{12}(t,t^{\prime})\left[a,\tilde{\rho}_{s}(t)\right],\\
\hat{O}_{s,22}(t,t^{\prime})=&x^{*}_{21}(t,t^{\prime})\left[a,\tilde{\rho}_{s}(t)\right].
\end{align}
Because the
operators~$\left\{a^{\dagger},\tilde{\rho}_{s}(t)\right\}$,~$\left[a^{\dagger},\tilde{\rho}_{s}(t)\right]$,
$\left\{a,\tilde{\rho}_{s}(t)\right\}$,~and~$\left[a,\tilde{\rho}_{s}(t)\right]$~are
arbitrary, it is straightforward to find out the equation of motion
for~$x_{mk}(t,t^{\prime})~(m,k=1,2)$~with Eqs.~(30)$-$(33). The
results are
\begin{align}
x_{11}(t,t^{\prime})=&\frac{1}{\sqrt{\hbar}}e^{-i\omega_{0}(t-t^{\prime})}
+\int_{t^{\prime}}^{t}dt_{1}
\int_{0}^{t_{1}}dt_{2}e^{-i\omega_{0}(t_{1}-t^{\prime})}\alpha_{1}(t_{1}-t_{2})x_{11}(t,t_{2}),\\
x_{12}(t,t^{\prime})=&\int_{t^{\prime}}^{t}dt_{1}
\int_{0}^{t_{1}}dt_{2}e^{-i\omega_{0}(t_{1}-t^{\prime})}\alpha_{1}(t_{1}-t_{2})x_{12}(t,t_{2})
\nonumber\\&-\int_{t^{\prime}}^{t}dt_{1}
\int_{0}^{t}dt_{2}e^{-i\omega_{0}(t_{1}-t^{\prime})}\alpha_{2}(t_{1}-t_{2})x_{21}(t,t_{2}),
\\
x_{21}(t,t^{\prime})=&\frac{1}{\sqrt{\hbar}}e^{-i\omega_{0}(t-t^{\prime})}
-\int_{t^{\prime}}^{t}dt_{1}
\int_{t_{1}}^{t}dt_{2}e^{-i\omega_{0}(t_{1}-t^{\prime})}\alpha^{*}_{1}(t_{2}-t_{1})x_{21}(t,t_{2}).
\end{align}
Whenever these coefficients are solved, the dissipative operators
Eqs.~(34)$-$(37) become available. Inserting into Eq.~(16), we
obtain the master equation,
\begin{align}
\frac{d\tilde{\rho}_{s}(t)}{dt}=&-iA_{1}(t)\left[a^{\dagger}a,\tilde{\rho}_{s}(t)\right]
+A_{2}(t)\left[2a\tilde{\rho}_{s}(t)a^{\dagger}-a^{\dagger}a\tilde{\rho}_{s}(t)-\tilde{\rho}_{s}(t)a^{\dagger}a\right]\nonumber\\&
+A_{3}(t)\left
[a^{\dagger}\tilde{\rho}_{s}(t)a+a\tilde{\rho}_{s}(t)a^{\dagger}-a^{\dagger}a\tilde{\rho}_{s}(t)-\tilde{\rho}_{s}(t)aa^{\dagger}\right],
\end{align}
where the coefficients~$A_{j}(t)~(j=1-3)$~are defined by
\begin{align}
A_{1}(t)=&\omega_{0}+\textup{Im}\left[\sqrt{\hbar}\int_{0}^{t}dt^{\prime}
\alpha^{*}_{1}(t-t^{\prime})x^{*}_{11}(t,t^{\prime})\right],\\
A_{2}(t)=&\textup{Re}\left[\sqrt{\hbar}\int_{0}^{t}dt^{\prime}\alpha^{*}_{1}(t-t^{\prime})x^{*}_{11}(t,t^{\prime})\right],\\
A_{3}(t)=&\textup{Re}\left[\sqrt{\hbar}\int_{0}^{t}dt^{\prime}\alpha^{*}_{2}(t-t^{\prime})
x^{*}_{21}(t,t^{\prime})\right]
-\textup{Re}\left[\sqrt{\hbar}\int_{0}^{t}dt^{\prime}\alpha^{*}_{1}(t-t^{\prime})y^{*}(t,t^{\prime})\right],
\end{align}
with
\begin{equation}
y^{*}(t,t^{\prime})=x^{*}_{11}(t,t^{\prime})+x^{*}_{12}(t,t^{\prime}).
\end{equation}
It is clear that~$A_{1}(t)$~is a frequency-renormalization
coefficient, $A_{2}(t)$~and~$A_{3}(t)$~are related to the
conventional dissipation and fluctuation coefficients, respectively.
In the following section, we will show the equivalence between our
derived master equation and that by Xiong {\em et al.} in terms of
path integral approach~\cite{Zhang}.
\section{Comparison with Known Results}
Resorting to the influence functional method developed by Feynman
and Vernon, Xiong {\em et al.} elaborated the derivation of the
master equation of optical cavity coupled to a heat
bath~\cite{Zhang}. For the case of a dissipative single-mode, their
result is of the same form as Eq.~(41) and the corresponding
coefficients read
\begin{align}
B_{1}(t)&=\omega_{0}+\textup{Im}\left[\sqrt{\hbar}\int_{0}^{t}dt^{\prime}\alpha^{*}_{1}(t-t^{\prime})\bar{x}_{11}(t,t^{\prime})\right],\\
B_{2}(t)&=\textup{Re}\left[\sqrt{\hbar}\int_{0}^{t}dt^{\prime}\alpha^{*}_{1}(t-t^{\prime})\bar{x}_{11}(t,t^{\prime})\right],\\
B_{3}(t)&=\textup{Re}\left[\sqrt{\hbar}\int_{0}^{t}dt^{\prime}\alpha^{*}_{2}(t-t^{\prime})\bar{x}_{21}(t,t^{\prime})\right]
-\textup{Re}\left[\sqrt{\hbar}\int_{0}^{t}dt^{\prime}\alpha^{*}_{1}(t-t^{\prime})\bar{y}(t,t^{\prime})\right].
\end{align}
The functions~$\bar{x}_{11}(t,t^{\prime})$,
$\bar{x}_{21}(t,t^{\prime})$, and~$\bar{y}(t,t^{\prime})$ are
defined by
\begin{align}
\bar{x}_{11}(t,t^{\prime})&=\frac{1}{\sqrt{\hbar}}u(t^{\prime})u^{-1}(t),\\
\bar{x}_{21}(t,t^{\prime})&=\frac{1}{\sqrt{\hbar}}u^{*}(t-t^{\prime}),\\
\bar{y}(t,t^{\prime})&=\frac{1}{\sqrt{\hbar}}\bar{u}^{*}(t^{\prime})-\frac{2}{\sqrt{\hbar}}\left[u(t^{\prime})u^{-1}(t)v(t)-v(t^{\prime})\right],
\end{align}
where~$u(t)$~and~$v(t)$~obey the following integro-differential
equations,
\begin{align}
\dot{u}(\tau)+i\omega_{0}u(\tau)+\int_{0}^{\tau}dt^{\prime}\alpha^{*}_{1}(\tau-t^{\prime})u(t^{\prime})&=0
\intertext{and}
\dot{v}(\tau)+i\omega_{0}v(\tau)+\int_{0}^{\tau}dt^{\prime}\alpha^{*}_{1}(\tau-t^{\prime})v(t^{\prime})
&=\frac{1}{2}\int_{0}^{t}dt^{\prime}\left[\alpha^{*}_{2}(\tau-t^{\prime})-\alpha^{*}_{1}(\tau-t^{\prime})\right]\bar{u}^{*}(t^{\prime})
\end{align}
with the initial conditions~$u(0)=1$,
$v(0)=0$,~and~$\bar{u}(\tau)\equiv u(t-\tau)$. To prove the
equivalence of the results obtained by two different methods we only
need to prove that~$A_{j}(t)=B_{j}(t)~(j=1-3)$, respectively. As
displayed in Eqs.~(42)$-$(44) and Eqs.~(46)$-$(48), all definite
integrals in the functions~$A_{j}(t)$~and~$B_{j}(t)$~are taken over
the same time range~$[0,t]$. Therefore, a sufficient condition
for~$A_{j}(t)=B_{j}(t)$~is that the corresponding integrands are
identical. Moreover, because these integrands consist of the
factors~$\alpha_{1}(t)$~and~$\alpha_{2}(t)$~that are dependent on
the specificity of the dissipation and can be arbitrary, one can
further simplify the problem as a proof of following
relations,\begin{align}
x^{*}_{11}(t,t^{\prime})&=\bar{x}_{11}(t,t^{\prime}),\\
x^{*}_{21}(t,t^{\prime})&=\bar{x}_{21}(t,t^{\prime}),\\
y^{*}(t,t^{\prime})&=\bar{y}(t,t^{\prime}).
\end{align}
\subsection{Proof of~$A_{1}(t)=B_{1}(t)$, $A_{2}(t)=B_{2}(t)$}
As clarified above, if Eq.~(54) holds,
then~$A_{1}(t)=B_{1}(t)$,~$A_{2}(t)=B_{2}(t)$. Note
that~$u(t)$~satisfies the linear integro-differential equation~(52)
and
that~$\bar{x}_{11}(t,t^{\prime})=u(t^{\prime})/(\sqrt{\hbar}u(t))$.
When the first argument~$t$~is fixed,
$\bar{x}_{11}(t,t^{\prime})$~can be seen as a function of the time
variable~$t^{\prime}$, which also obeys Eq.~(52), namely,
\begin{equation}
\frac{\partial}{\partial
t^{\prime}}\bar{x}_{11}(t,t^{\prime})+i\omega_{0}\bar{x}_{11}(t,t^{\prime})+\int_{0}^{t^{\prime}}dt_{1}\alpha^{*}_{1}(t^{\prime}-t_{1})\bar{x}(t,t_{1})=0.
\end{equation}
Return to the integral equation of~$x_{11}(t,t^{\prime})$, Eq.~(38).
Calculating the time derivative with respect to~$t^{\prime}$~and
taking the operation of complex conjugation on both sides of
Eq.~(38), one obtains for~$x^{*}_{11}(t,t^{\prime})$~the same
equation as Eq.~(57). Also, the initial condition for these
equations are the same, namely,~$
x_{11}^{*}(t,t^{\prime})|_{t^{\prime}=t}=\bar{x}_{11}(t,t^{\prime})|_{t^{\prime}=t}=1/\sqrt{\hbar}$.
Therefore, $A_{1}(t)=B_{1}(t)$~and~$A_{2}(t)=B_{2}(t)$~are proved.
\subsection{Proof of $A_{3}(t)=B_{3}(t)$}
One only needs to demonstrate that Eqs.~(55)~and~(56) hold. A
straightforward algebra shows that~$x_{21}(t,t^{\prime})$~is
time-translation invariant, i.e.,
$x_{21}(t,t^{\prime})=x_{21}(t+\lambda,t^{\prime}+\lambda)$,
where~$\lambda$~is a constant. It means that~$x_{21}$~is a function
of the time difference~$t-t^{\prime}$,
$x_{21}(t,t^{\prime})=x_{21}(t-t^{\prime})$. As a result, Eq.~(40)
can be simplified as
\begin{equation}
x_{21}(s)=\frac{1}{\sqrt{\hbar}}e^{-i\omega_{0}s}-\int_{0}^{s}dt_{1}\int_{t_{1}}^{s}dt_{2}e^{-i\omega_{0}t_{1}}
\alpha_{1}^{*}(t_{2}-t_{1})x_{21}(s-t_{2}).
\end{equation}
Taking the first-order derivation with respect to the
argument~$s$~and the complex conjugation, one obtains
\begin{equation}
\frac{d}{d
s}x_{21}^{*}(s)=i\omega_{0}x_{21}^{*}(s)-\int_{0}^{s}dt_{1}\alpha^{*}_{1}(t_{1}-s)x_{21}^{*}(t_{1}),
\end{equation}
subjected to the initial
condition~$x^{*}_{21}(s)|_{s=0}=1/\sqrt{\hbar}$. By definition
Eq.~(50), the function~$\bar{x}_{21}(t,t^{\prime})$~is only
dependent on the time difference~$s=t-t^{\prime}$. Taking the
operation of complex conjugation on both sides of Eq.~(52) leads to
the equation which is the same as Eq.~(59). Besides,
$\bar{x}_{21}(s)|_{s=0}=1/\sqrt{\hbar}=x_{21}^{*}(s)|_{s=0}$.
Therefore,~$x_{21}^{*}(t,t^{\prime})=\bar{x}_{21}(t,t^{\prime})$~does
hold.
\par
By definition Eq.~(51) and with the help of Eqs.~(52) and (53), we
find the that~$\bar{y}(t,t^{\prime})$~satisfies
\begin{align}
\frac{\partial}{\partial t^{\prime}} \bar{y}(t,t^{\prime})
&=-i\omega_{0}\bar{y}(t,t^{\prime})-\int_{0}^{t^{\prime}}dt_{1}\alpha^{*}_{1}(t^{\prime}-t_{1})\bar{y}(t,t_{1})+\int_{0}^{t}dt_{1}
\alpha^{*}_{2}(t^{\prime}-t_{1})\bar{x}_{21}(t,t_{1}).
\end{align}
The same equation can be obtained for~$y^{*}(t,t^{\prime})$~from
Eqs.~(38) and (39). Moreover,
$\bar{y}(t,t^{\prime})|_{t^{\prime}=t}=y^{*}(t,t^{\prime})|_{t^{\prime}=t}=1/\sqrt{\hbar}$.
Therefore, one
proves~$\bar{y}(t,t^{\prime})=y^{*}(t,t^{\prime})$~and as a
result,~$A_{3}(t)=B_{3}(t)$. We have therefore demonstrated that the
master equation Eq.~(41) resulting from stochastic description is
identical with that derived with influence functional
method~\cite{Zhang}. \par Some remarks on the calculation of the
coefficients of the master equation are in order. As discussed
above, our procedure provides a set of integral equations, while
Xiong {\em et al}.\cite{Zhang} introduce an integro-differential
equation or the equation of the related Green's function. It is
straightforward to numerically independent both of the two schemes
to determine the coefficients. Although we prove that these two
frameworks give the identical results, their numerical performance
might be different. Because the computational scaling for solving
the integral equation is less favorable than solving the
corresponding differential equation, the Green's method is preferred
in practice.
\section{Driven cavity dynamics}
Let us consider the cavity dynamics in the presence of a
time-dependent external field~$\epsilon(t)$. Now the Hamiltonian of
the system
reads~$\hat{H}_{s}(t)=\hbar\omega_{0}a^{\dagger}a+\epsilon(t)\left(a+a^{\dagger}\right)$.
The master equation can be derived along the same line discussed in
Sec.~\Rmnum{3}. Although the external field only directly acts on
the cavity system, and does not change the bath-induced stochastic
fields, it does interfere with the bath during the evolution of the
system. This effect is reflected in the change of dissipative
operators. Starting with Eqs.~(24)$-$(27), we solve the required
propagators and take the stochastic averaging to obtain
\begin{align*}
\hat{O}_{s,11}(t,t^{\prime})=&\frac{1}{\sqrt{\hbar}}\left\{a^{\dagger},\tilde{\rho}_{s}(t)\right\}e^{-i\omega_{0}
(t-t^{\prime})}+\int_{t^{\prime}}^{t}dt_{1}\int_{0}^{t_{1}}dt_{2}e^{-i\omega_{0}(t_{1}-t^{\prime})}
\alpha_{1}(t_{1}-t_{2})\hat{O}_{s,11}(t,t_{2})
\nonumber\\&-\int_{t^{\prime}}^{t}dt_{1}\int_{0}^{t}dt_{2}e^{-i\omega_{0}(t_{1}-t^{\prime})}
\alpha_{2}(t_{1}-t_{2})\hat{O}_{s,12}(t,t_{2})
-\frac{2i}{\hbar\sqrt{\hbar}}\int_{t^{\prime}}^{t}dt_{1}e^{-i\omega_{0}(t_{1}-t^{\prime})}\epsilon(t_{1})\tilde{\rho}_{s}(t),\\
\hat{O}_{s,12}(t,t^{\prime})=&\frac{1}{\sqrt{\hbar}}\left[a^{\dagger},\tilde{\rho}_{s}(t)\right]e^{-i\omega_{0}
(t-t^{\prime})}-\int_{t^{\prime}}^{t}dt_{1}
\int_{t_{1}}^{t}dt_{2}e^{-i\omega_{0}(t_{1}-t^{\prime})}\alpha^{*}_{1}(t_{2}-t_{1})\hat{O}_{s,12}(t,t_{2}),\\
\hat{O}_{s,21}(t,t^{\prime})=&\frac{1}{\sqrt{\hbar}}
\left\{a,\tilde{\rho}_{s}(t)\right\}e^{i\omega_{0}
(t-t^{\prime})}+\int_{t^{\prime}}^{t}dt_{1}\int_{0}^{t_{1}}dt_{2}e^{i\omega_{0}(t_{1}-t^{\prime})}
\alpha^{*}_{1}(t_{1}-t_{2})\hat{O}_{s,21}(t,t_{2})
\nonumber\\&+\int_{t^{\prime}}^{t}dt_{1}\int_{0}^{t}dt_{2}e^{i\omega_{0}(t_{1}-t^{\prime})}\alpha_{2}(t_{2}-t_{1})
\hat{O}_{s,22}(t,t_{2})+\frac{2i}{\hbar\sqrt{\hbar}}\int_{t^{\prime}}^{t}dt_{1}e^{i\omega_{0}(t_{1}-t^{\prime})}\epsilon(t_{1})\tilde{\rho}_{s}(t),
\\
\hat{O}_{s,22}(t,t^{\prime})=&\frac{1}{\sqrt{\hbar}}\left[a,\tilde{\rho}_{s}(t)\right]e^{i\omega_{0}
(t-t^{\prime})}-\int_{t^{\prime}}^{t}dt_{1}
\int_{t_{1}}^{t}dt_{2}e^{i\omega_{0}(t_{1}-t^{\prime})}\alpha_{1}(t_{2}-t_{1})\hat{O}_{s,22}(t,t_{2}).\end{align*}
We use the same reasoning as that in Sec.~\Rmnum{3} to obtain
\begin{align*}
\hat{O}_{s,11}(t,t^{\prime})&=x_{11}(t,t^{\prime})\left\{a^{\dagger},\tilde{\rho}_{s}(t)\right\}
+x_{12}(t,t^{\prime})\left[a^{\dagger},\tilde{\rho}_{s}(t)\right]
+x_{13}(t,t^{\prime})\tilde{\rho}_{s}(t),\\
\hat{O}_{s,12}(t,t^{\prime})&=x_{21}(t,t^{\prime})\left[a^{\dagger},\tilde{\rho}_{s}(t)\right],\\
\hat{O}_{s,21}(t,t^{\prime})&=x^{*}_{11}(t,t^{\prime})\left\{a,\tilde{\rho}_{s}(t)\right\}-x^{*}_{12}(t,t^{\prime})\left[a,\tilde{\rho}_{s}(t)\right]
+x^{*}_{13}(t,t^{\prime})\tilde{\rho}_{s}(t),\\
\hat{O}_{s,22}(t,t^{\prime})&=x^{*}_{21}(t,t^{\prime})\left[a,\tilde{\rho}_{s}(t)\right],\end{align*}
where all coefficients except~$x_{13}(t,t^{\prime})$~are the same as
that of the undriven case [Eqs.~(38)$-$(40)]. The additional new
function is defined by
\begin{align*}
x_{13}(t,t^{\prime})&=-\frac{2i}{\hbar\sqrt{\hbar}}\int_{t^{\prime}}^{t}dt_{1}e^{-i\omega_{0}(t_{1}-t^{\prime})}\epsilon(t_{1})
+\int_{t^{\prime}}^{t}dt_{1}
\int_{0}^{t_{1}}dt_{2}e^{-i\omega_{0}(t_{1}-t^{\prime})}\alpha_{1}(t_{1}-t_{2})x_{13}(t,t_{2}),
\end{align*}
which is linearly dependent on the external driving field
$\epsilon(t)$.
\par With these expressions the master equation now reads
\begin{align}
\frac{d\tilde{\rho}_{s}(t)}{dt}=&\left[-iA_{1}(t)a^{\dagger}a+C(t)a+D(t)a^{\dagger},\tilde{\rho}_{s}(t)\right]
+A_{2}(t)\left[2a\tilde{\rho}_{s}(t)a^{\dagger}-a^{\dagger}a\tilde{\rho}_{s}(t)-\tilde{\rho}_{s}(t)a^{\dagger}a\right]\nonumber\\&
+A_{3}(t)
\left[a^{\dagger}\tilde{\rho}_{s}(t)a+a\tilde{\rho}_{s}(t)a^{\dagger}-a^{\dagger}a\tilde{\rho}_{s}(t)-\tilde{\rho}_{s}(t)aa^{\dagger}\right],
\end{align}
where
\begin{equation*}
C(t)=-\frac{i}{\hbar}\epsilon(t)
+\frac{\sqrt{\hbar}}{2}\int_{0}^{t}dt^{\prime}\alpha_{1}(t-t^{\prime})x_{13}(t,t^{\prime})\end{equation*}
and
\begin{equation*}
D(t)=-\frac{i}{\hbar}\epsilon(t)
-\frac{\sqrt{\hbar}}{2}\int_{0}^{t}dt^{\prime}\alpha^{*}_{1}(t-t^{\prime})x^{*}_{13}(t,t^{\prime}).
\end{equation*}
Here, the coefficients~$A_{j}(t)~(j=1-3)$~are the same as that of
the undriven case, which satisfy Eqs.~(42)$-$(44). It becomes clear
that there are effects of the external field on the system, one is
the direct interaction and the other results in the very interplay
between the driving field and dissipation. As a consequence, the
external field can be applied to {\em control} dissipation, or via
versa, dissipation can be used to modulate the external field.
\section{Master equation for two-state atoms in Vacuum}
The spontaneous decay of a two-state atom coupled to a vacuum is
described by the Hamiltonian Eq.~(1)
with~$\hat{H}_{s}=-\hbar\omega_{0}\sigma_{z}/2$,
$\hat{f}_{1}=\sigma^{-}$, and~$\hat{f}_{2}=\sigma^{+}$,
where~$\sigma_{z}$~is the pauli matrix,
and~$\sigma^{+}$~and~$\sigma^{-}$~are the raising and lowering
operators. They satisfy the commutation
relations~$\left[\sigma^{+},\sigma^{-}\right]=\sigma_{z}$,
$\left[\sigma^{+},\sigma_{z}\right]=-2\sigma^{+}$,
and~$\left[\sigma^{-},\sigma_{z}\right]=2\sigma^{-}$. This damped
two-state model might provide fundamental understanding of
decoherence and other features of the dynamics of a qubit coupled to
a heat bath. It is no wonder that its master equation has been
derived and explored by several authors with diversified theoretical
methods. For instance, Garraway developed a pseudomode technique to
solve the dynamics~\cite{Garraway}. Through the solution of the
Schr\"{o}dinger equation for the entire system, Breuer and coworkers
worked out the reduced density matrix and thereby proposed a simple
derivation of the corresponding master equation by a brute force
calculation of the derivative with respect to time~\cite{Breure,
H.P.Breuer}. They also developed a stochastic wave function approach
to simulate the dynamics~\cite{Vacchini}. Strunz {\em et al.}
proposed a different stochastic Schr\"{o}dinger function method to
solve the dissipative dynamics of the model~\cite{Strunz}. The
reduced density matrix resulting from the Schr\"{o}dinger equation
was also exploited by Vacchini and coworkers who recently showed how
to generate the exact master equations corresponding to the
time-convolutionless form and to the Nakajima-Zwanzig non-Markovian
form~\cite{Smirne}. In the first paper on the stochastic description
of quantum dissipative systems, one of the authors also demonstrated
how to derive the master equation from the stochastic equation of
motion~\cite{J.Shao}. His method is based on self-consistency of an
ansatz related to a stochastic average and the derivation was not
expounded in the paper~\cite{J.Shao}.\par It seems that all the
derivations in the literature are not direct and straightforward
within one theoretical framework. We will show the stochastic
description does offer a good pass to the master equation from the
equation for the random density matrix for the system. Because the
bath is the vacuum field, the temperature is zero. As a
result,~$\coth\left[\hbar\omega/(2k_{B}T)\right]=\textup{I}$~and the
response functions derived by Eqs.~(13) and (14) become
identical,~$\alpha_{1}(t)=\alpha_{2}(t)\equiv\alpha(t)$. Therefore,
the bath-induced stochastic fields determined by Eqs.~(11) and (12)
become
\begin{equation}
\bar{g}_{1}(t)=\frac{i\hbar\sqrt{\hbar}}{2}\int_{0}^{t}dt^{\prime}\alpha(t-t^{\prime})
\left[-i\nu_{12}(t^{\prime})+\nu_{22}(t^{\prime})+i\nu_{32}(t^{\prime})-\nu_{42}(t^{\prime})\right]
\end{equation}
and
\begin{equation}
\bar{g}_{2}(t)=-\frac{i\hbar\sqrt{\hbar}}{2}\int_{0}^{t}dt^{\prime}\alpha^{*}(t-t^{\prime})
\left[i\nu_{11}(t^{\prime})+\nu_{21}(t^{\prime})+i\nu_{31}(t^{\prime})+\nu_{41}(t^{\prime})\right].
\end{equation}

The formal solution of the random density matrix of the system is
still given by Eq.~(21) where the forward and backward
propagators~$U_{1}(t,0)$~and~$U_{2}(0,t)$~are ruled by the
corresponding Hamiltonians Eqs.~(22) and (23)
with~$\hat{f}_{1}=\sigma^{-}$~and~$\hat{f}_{2}=\sigma^{+}$.
Apparently, there are six complex Gaussian
fields,~$\bar{g}_{1}(t)$,~$\bar{g}_{2}(t)$,~$\eta_{11}^{*}(t)$,~$\eta_{12}^{*}(t)$,~$\eta_{21}^{*}(t)$,
and~$\eta_{22}^{*}(t)$~involving in the dynamics. Note that all of
the six Gaussian noises have zero means and null autocovariances.
The average of a stochastic process generated, therefore, is fully
determined by their non-vanishing cross-covariances.
Given~$\bar{g}_{1}(t)$~and~$\bar{g}_{2}(t)$~by Eqs. (62) and (63),
however, one can readily check that the white
noises~$\eta_{12}^{*}(t)$~and~$\eta_{21}^{*}(t)$~are not correlated
with other four and do not have any influence on the averaged
dynamics. Therefore,~$\eta_{12}^{*}(t)$~and~$\eta_{21}^{*}(t)$~can
be safely omitted when calculating the reduced density matrix.

To derive the master equation, we
insert~$\bar{g}_{1}(t)$~and~$\bar{g}_{2}(t)$ into Eq.~(2) and take
stochastic averaging to obtain
\begin{align}
i\hbar\frac{\partial \tilde{\rho}_{s}(t)}{\partial
t}=&\left[\hat{H}_{s},\tilde{\rho}_{s}(t)\right]+\frac{i\hbar\sqrt{\hbar}}{2}
\left[\sigma^{-},\int_{0}^{t}dt^{\prime}\alpha(t-t^{\prime})\hat{O}_{s,1}(t,t^{\prime})
\right]
\nonumber\\&-\frac{i\hbar\sqrt{\hbar}}{2}\left[\sigma^{+},\int_{0}^{t}dt^{\prime}\alpha^{*}(t-t^{\prime})
\hat{O}_{s,2}(t,t^{\prime})\right],\end{align} where the dissipative
operators are
\begin{align}
\hat{O}_{s,1}(t,t^{\prime})&=M\left\{-i\frac{\delta\rho_{s}(t)}{\delta\nu_{12}(t^{\prime})}
+\frac{\delta\rho_{s}(t)}{\delta\nu_{22}(t^{\prime})}+i\frac{\delta\rho_{s}(t)}{\delta\nu_{32}(t^{\prime})}-\frac{\delta\rho_{s}(t)}{\delta\nu_{42}(t^{\prime})}\right\}
\nonumber\\&=\frac{2}{\sqrt{\hbar}}M\left\{\rho_{s}(t)\sigma_{2}^{+}(t,t^{\prime})\right\}
\end{align}
and
\begin{align}
\hat{O}_{s,2}(t,t^{\prime})&=M\left\{i\frac{\delta\rho_{s}(t)}{\delta\nu_{11}(t^{\prime})}+\frac{\delta\rho_{s}(t)}{\delta\nu_{21}(t^{\prime})}
+i\frac{\delta\rho_{s}(t)}{\delta\nu_{31}(t^{\prime})}+\frac{\delta\rho_{s}(t)}{\delta\nu_{41}(t^{\prime})}\right\}
\nonumber\\&=\frac{2}{\sqrt{\hbar}}M\left\{\sigma_{1}^{-}(t,t^{\prime})\rho_{s}(t)\right\},
\end{align}
with~$\sigma^{\pm}_{1,2}(t,t^{\prime})=U_{1,2}(t,t^{\prime})\sigma^{\pm}U_{1,2}(t^{\prime},t)$.
We like to stress that the derivation up to now is parallel to that
illuminated in Sec.~\Rmnum{3}. Now we need to find the explicit
expressions
of~$M\left\{\rho_{s}(t)\sigma_{2}^{+}(t,t^{\prime})\right\}$~and
$M\left\{\sigma_{1}^{-}(t,t^{\prime})\rho_{s}(t)\right\}$~in terms
of~$\tilde{\rho}_{s}(t)$~and other known operators of the system. To
this end, we consider their derivatives with respect
to~$t^{\prime}$,
\begin{align}
\frac{\partial}{\partial
t^{\prime}}M\left\{\rho_{s}(t)\sigma_{2}^{+}(t,t^{\prime})\right\}
=&-i\omega_{0}M\left\{\rho_{s}(t)\sigma_{2}^{+}(t,t^{\prime})\right\}\nonumber\\
&+\int_{0}^{t^{\prime}}dt_{1}\alpha(t^{\prime}-t_{1})M\left\{\rho_{s}(t)\sigma^{+}_{2}(t,t_{1})U_{2}(t,t^{\prime})\sigma_{z}U_{2}(t^{\prime},t)\right\}
\end{align}
and
\begin{align}
\frac{\partial}{\partial
t^{\prime}}M\left\{\sigma^{-}_{1}(t,t^{\prime})\rho_{s}(t)\right\}
=&i\omega_{0}M\left\{\sigma^{-}_{1}(t,t^{\prime})\rho_{s}(t)\right\}
\nonumber\\&+\int_{0}^{t^{\prime}}dt_{1}\alpha^{*}(t^{\prime}-t_{1})M\left\{
U_{1}(t,t^{\prime})\sigma_{z}U_{1}(t^{\prime},t)\sigma^{-}_{1}(t,t_{1})\rho_{s}(t)\right\}.
\end{align}
By virtue of~$\sigma_{z}=2\sigma^{+}\sigma^{-}-\textup{I}$, the two
equations can be converted to
\begin{align}
\frac{\partial}{\partial
t^{\prime}}M\left\{\rho_{s}(t)\sigma_{2}^{+}(t,t^{\prime})\right\}
=&-i\omega_{0}M\left\{\rho_{s}(t)\sigma_{2}^{+}(t,t^{\prime})\right\}+2\int_{0}^{t^{\prime}}dt_{1}\alpha(t^{\prime}-t_{1})M\left\{
\widehat{X}_{1}(t,t_{1},t^{\prime})\right\}\nonumber\\
&-\int_{0}^{t^{\prime}}dt_{1}\alpha(t^{\prime}-t_{1})
M\left\{\rho_{s}(t)\sigma_{2}^{+}(t,t_{1})\right\}
\end{align}
and \begin{align} \frac{\partial}{\partial
t^{\prime}}M\left\{\sigma_{1}^{-}(t,t^{\prime})\rho_{s}(t)\right\}
=&
i\omega_{0}M\left\{\sigma_{1}^{-}(t,t^{\prime})\rho_{s}(t)\right\}+2\int_{0}^{t^{\prime}}dt_{1}\alpha^{*}(t^{\prime}-t_{1})M\left\{
\widehat{X}_{2}(t,t_{1},t^{\prime})\right\}\nonumber\\&
-\int_{0}^{t^{\prime}}dt_{1}\alpha^{*}(t^{\prime}-t_{1})M\left\{\sigma_{1}^{-}(t,t_{1})\rho_{s}(t)\right\},
\end{align}
where \begin{align*}
\widehat{X}_{1}(t,t_{1},t^{\prime})&=\rho_{s}(t)
\sigma_{2}^{+}(t,t_{1})\bar{\sigma}_{2}(t,t^{\prime})\intertext{and}
\widehat{X}_{2}(t,t_{1},t^{\prime})&=\bar{\sigma}_{1}(t,t^{\prime})\sigma_{1}^{-}(t,t_{1})\rho_{s}(t),\end{align*}
with~$\bar{\sigma}_{1,2}(t,t^{\prime})=U_{1,2}(t,t^{\prime})\sigma^{+}\sigma^{-}U_{1,2}(t^{\prime},t)$.
\par To find closed equations
for~$M\left\{\rho_{s}(t)\sigma_{2}^{+}(t,t^{\prime})\right\}$~and~$M\left\{\sigma_{1}^{-}(t,t^{\prime})\rho_{s}(t)\right\}$,
therefore, we should
evaluate~$M\left\{\widehat{X}_{1}(t,t_{1},t^{\prime})\right\}$~and~$M\left\{\widehat{X}_{2}(t,t_{1},t^{\prime})\right\}$.
When the first argument~$t$~is fixed,
$M\left\{\widehat{X}_{1}(t,t_{1},t^{\prime})\right\}$~and~$M\left\{\widehat{X}_{2}(t,t_{1},t^{\prime})\right\}$~can
be taken as the functions of~$t_{1}$~and~$t^{\prime}$. For brevity,
the argument~$t$~for
functions~$\widehat{X}_{1}$~and~$\widehat{X}_{2}$~will not be
written. On taking their derivatives with respect to~$t_{1}$, we
obtain
\begin{align}
\frac{\partial}{\partial
t_{1}}M\left\{\widehat{X}_{1}(t_{1},t^{\prime})\right\}=&-i\omega_{0}M\left\{\widehat{X}_{1}(t_{1},t^{\prime})\right\}
+\int_{0}^{t_{1}}dt_{2}\alpha(t_{1}-t_{2})M\left\{2\widehat{X}_{1}(t_{2},t_{1})\bar{\sigma}_{2}(t,t^{\prime})-\widehat{X}_{1}(t_{2},t^{\prime})\right\}
\end{align}
and
\begin{align}
\frac{\partial}{\partial
t_{1}}M\left\{\widehat{X}_{2}(t_{1},t^{\prime})\right\}=&i\omega_{0}M\left\{\widehat{X}_{2}(t_{1},t^{\prime})\right\}
+\int_{0}^{t_{1}}dt_{2}\alpha^{*}(t_{1}-t_{2})M\left\{2\bar{\sigma}_{1}(t,t^{\prime})\widehat{X}_{2}(t_{2},t_{1})-\widehat{X}_{2}(t_{2},t^{\prime})\right\}.
\end{align}
We like to point out that the solutions
for~$\widehat{X}_{1}(t_{1},t^{\prime})$~and~$\widehat{X}_{2}(t_{1},t^{\prime})$~can
be many as long as their {\em stochastic averages} satisfy Eqs.~(71)
and (72). Because any solutions are sufficient for our purpose, we
only consider the simple ones determined by
\begin{align}
\frac{\partial}{\partial
t_{1}}\widehat{X}_{1}(t_{1},t^{\prime})=&-i\omega_{0}\widehat{X}_{1}(t_{1},t^{\prime})+\int_{0}^{t_{1}}dt_{2}\alpha(t_{1}-t_{2})
\left[2\widehat{X}_{1}(t_{2},t_{1})\bar{\sigma}_{2}(t,t^{\prime})-\widehat{X}_{1}(t_{2},t^{\prime})\right]
\end{align}
and
\begin{align}
\frac{\partial}{\partial
t_{1}}\widehat{X}_{2}(t_{1},t^{\prime})=&i\omega_{0}\widehat{X}_{2}(t_{1},t^{\prime})
+\int_{0}^{t_{1}}dt_{2}\alpha^{*}(t_{1}-t_{2})\left[2\bar{\sigma}_{1}(t,t^{\prime})\widehat{X}_{2}(t_{2},t_{1})-\widehat{X}_{2}(t_{2},t^{\prime})\right],
\end{align}
with the initial conditions
$\widehat{X}_{1}(t_{1},t^{\prime})|_{t_{1}=t^{\prime}}=0$~and~$\widehat{X}_{2}(t_{1},t^{\prime})|_{t_{1}=t^{\prime}}=0$.

As a result, we
obtain~$\widehat{X}_{1}(t_{1},t^{\prime})=0$~and~$\widehat{X}_{2}(t_{1},t^{\prime})=0$.
Then Eqs.~(69) and (70) become
\begin{align}
\frac{\partial}{\partial
t^{\prime}}M\left\{\rho_{s}(t)\sigma_{2}^{+}(t,t^{\prime})\right\}
&=-i\omega_{0}M\left\{\rho_{s}(t)\sigma_{2}^{+}(t,t^{\prime})\right\}
-\int_{0}^{t^{\prime}}dt_{1}\alpha(t^{\prime}-t_{1})M\left\{\rho_{s}(t)
\sigma_{2}^{+}(t,t_{1})\right\} \intertext{and}
\frac{\partial}{\partial
t^{\prime}}M\left\{\sigma_{1}^{-}(t,t^{\prime})\rho_{s}(t)\right\}
&=i\omega_{0}M\left\{\sigma_{1}^{-}(t,t^{\prime})\rho_{s}(t)\right\}-\int_{0}^{t^{\prime}}dt_{1}\alpha^{*}(t^{\prime}-t_{1})M\left\{\sigma_{1}^{-}(t,t_{1})\rho_{s}(t)\right\}.
\end{align}
They are integrated over time~$t^{\prime}$, namely,
\begin{align}
M\left\{\rho_{s}(t)\sigma_{2}^{+}(t,t^{\prime})\right\}
=&e^{-i\omega_{0}\left(t^{\prime}-t\right)}\tilde{\rho}_{s}(t)\sigma^{+}
+\int_{t^{\prime}}^{t}dt_{1}\int_{0}^{t_{1}}dt_{2}e^{-i\omega_{0}\left(t^{\prime}-t_{1}\right)}\alpha(t_{1}-t_{2})M\left\{\rho_{s}(t)
\sigma_{2}^{+}(t,t_{2})\right\}
\end{align}
and
\begin{align}
M\left\{\sigma_{1}^{-}(t,t^{\prime})\rho_{s}(t)\right\}
=&e^{i\omega_{0}\left(t^{\prime}-t\right)}\sigma^{-}\tilde{\rho}_{s}(t)
+\int_{t^{\prime}}^{t}dt_{1}\int_{0}^{t_{1}}dt_{2}e^{i\omega_{0}\left(t^{\prime}-t_{1}\right)}\alpha^{*}(t_{1}-t_{2})M\left\{
\sigma_{1}^{-}(t,t_{2})\rho_{s}(t)\right\}.
\end{align}
Note
that~$M\left\{\rho_{s}(t)\sigma_{2}^{+}(t,t^{\prime})\right\}^{\dagger}=M\left\{\sigma_{1}^{-}(t,t^{\prime})\rho_{s}(t)\right\}$.
By iteration, one can find
that~$M\left\{\rho_{s}(t)\sigma_{2}^{+}(t,t^{\prime})\right\}$
and~$M\left\{\sigma_{1}^{-}(t,t^{\prime})\rho_{s}(t)\right\}$~posses
the following forms,
\begin{equation}
M\left\{\rho_{s}(t)\sigma_{2}^{+}(t,t^{\prime})\right\}
=x(t,t^{\prime})\tilde{\rho}_{s}(t)\sigma^{+}
\end{equation}
and
\begin{equation}
M\left\{\sigma_{1}^{-}(t,t^{\prime})\rho_{s}(t)\right\}
=x^{*}(t,t^{\prime})\sigma^{-}\tilde{\rho}_{s}(t).
\end{equation}
Because the
operators~$\tilde{\rho}_{s}(t)\sigma^{+}$~and~$\sigma^{-}\tilde{\rho_{s}}(t)$~are
arbitrary, the coefficient~$x(t,t^{\prime})$~is determined by
Eq.~(77), which obeys
\begin{align}
x(t,t^{\prime})&=
e^{-i\omega_{0}\left(t^{\prime}-t\right)}+\int_{t^{\prime}}^{t}dt_{1}\int_{0}^{t_{1}}dt_{2}e^{-i\omega_{0}\left(t^{\prime}-t_{1}\right)}\alpha(t_{1}-t_{2})x(t,t_{2}).
\end{align} With the explicit
expressions
of~$M\left\{\rho_{s}(t)\sigma_{2}^{+}(t,t^{\prime})\right\}$~and
$M\left\{\sigma_{1}^{-}(t,t^{\prime})\rho_{s}(t)\right\}$,~Eq.~(64)
immediately becomes the resulting master equation. For the
spontaneous decay of a two-state atom it reads
\begin{align}
\frac{d\tilde{\rho}_{s}(t)}{dt}=&
-\frac{i}{\hbar}\left[\hat{H}_{s},\tilde{\rho}_{s}(t)\right]-i\frac{S(t)}{2}\left[\sigma^{+}\sigma^{-},\tilde{\rho}_{s}(t)\right]\nonumber\\&+R(t)\left[\sigma^{-}\tilde{\rho}_{s}(t)\sigma^{+}
-\frac{1}{2}\sigma^{+}\sigma^{-}\tilde{\rho}_{s}(t)-\frac{1}{2}\tilde{\rho}_{s}(t)\sigma^{+}\sigma^{-}\right],
\end{align}
where~$S(t)$~and~$R(t)$~are the time-dependent coefficients for the
descriptions of a frequency shift and a decay rate, respectively.
Their expressions are
\begin{align*}
R(t)=&2\textup{Re}\left[\int_{0}^{t}dt^{\prime}\alpha^{*}(t-t^{\prime})x^{*}(t,t^{\prime})\right]\intertext{and}
S(t)=&2\textup{Im}\left[\int_{0}^{t}dt^{\prime}\alpha^{*}(t-t^{\prime})x^{*}(t,t^{\prime})\right].
\end{align*}
\section{Conclusion}
The main goal of investigating dissipative systems is to solve their
properties, in particular, to reveal the dissipative effect on their
dynamics or Brownian motion. From the system plus environment model,
we have shown~\cite{J.Shao} that the coupling to the environment can
be rigorously mapped into stochastic fields and thereby provided a
microscopic description of the Brownian motion. The resulting
equation of motion for the density operator is a stochastic
Liouville equation and the statistical average of the solution gives
the reduced density matrix, the key quantity defining the system.
Like the classical counterpart, the Langevin equation, the
stochastic Liouville equation offers a convenient way for the
numerical simulation of quantum Brownian dynamics, however, its
efficiency is seriously limited due to the slow convergence of
stochastic averaging~\cite{Shao1,Shao2,Shao3}. It is therefore
desirable to derive the equation of motion for the reduced density
operator or the master equation if it exists, given the stochastic
Liouville equation. A general procedure was suggested
in~\cite{J.Shao} and the detailed derivation of the master equation
for the dissipative harmonic oscillator was presented in~\cite{Li}.
This paper tackles the dissipative dynamics of quantum optics in the
same light. \par We first worked out the bath-induced stochastic
fields comprising two terms with the rotating-wave-approximation
type interaction and then showed how to determine the ``dissipation
operators" for a single cavity mode. Similar to the case of the
dissipative harmonic oscillator described by the Caldeira-Leggett
model, the coefficients of the master equation for single cavity
mode are determined by a set of integral equations. It is shown that
our result is identical to that derived by virtue of path integral
technique~\cite{Zhang}. The master equation of a dissipative cavity
mode at a driving field was also derived and the display between the
dissipation and the driving field was pointed out. To show that the
stochastic formulation is a systematic method for treating
dissipative dynamics in quantum optics, we finally explained how to
acquire the master equation for the spontaneous decay of two-state
atoms coupled to the vacuum field. For solving the master equation,
because the integral equation is time-nonlocal, it would be better
to transform it into a differential one for a favorable numerical
implementation, if such a transformation is available. \par There
are still many interesting questions in the stochastic formulation
of dissipation. A related one to the derivation of the master
equation is for what kinds of system and couplings such an equation
exists. Notwithstanding, as the quantum dissipation becomes an
important and subtle issue and attracts more and more attention in
the community of quantum optics and quantum information, it is
expected that the stochastic description will be a powerful tool in
either theoretical analysis or numerical simulations.
\section*{Acknowledgments}
This work is supported by the National Natural Science Foundation of
China (No. 91027013) and the 973 program of the Ministry of Science
and Technology of China (2011CB808502).

\end{document}